\documentclass[12pt]{article}

\usepackage{array}
\usepackage{epsfig}
\usepackage{amsmath}
\usepackage{cite}
\usepackage{amssymb}
\usepackage{graphics,graphpap}

\usepackage[active]{srcltx}

\newcommand{\nsl}{\mbox{$n$\hspace{-0.5em}\raisebox{0.1ex}{$/$}}}
\newcommand{\nbarsl}{\mbox{$\bar{n}$\hspace{-0.5em}\raisebox{0.1ex}{$/$}}}
\newcommand{\vsl}{\mbox{$v$\hspace{-0.5em}\raisebox{0.1ex}{$/$}}}

\def\s#1{\setbox0=\hbox{$#1$}%
  \rlap{\ifdim\wd0>.7em\kern.22\wd0\else\kern.1\wd0\fi /}#1}

\def\cusp{\hbox{\tiny cusp}}
\def\MSbar{\hbox{\tiny ${\overline{\rm MS}}$}}
\catcode`\@=11 
\def\lsim{\mathrel{\mathpalette\@versim<}}
\def\gsim{\mathrel{\mathpalette\@versim>}}
\def\@versim#1#2{\vcenter{\offinterlineskip
        \ialign{$\m@th#1\hfil##\hfil$\crcr#2\crcr\sim\crcr } }}
\catcode`\@=12 


\begin{document}

\begin{titlepage}
\begin{flushright}
\begin{tabular}{l} 
IPPP/08/18\\
DCPT/08/36\\
Edinburgh 2008/15\\
\end{tabular}
\end{flushright}
\vskip1.5cm
\begin{center}
{\Large \bf \boldmath
Distribution Amplitudes of the $\Lambda_b$ Baryon in QCD}
\vskip1.3cm 
{\sc
Patricia Ball\footnote{Patricia.Ball@durham.ac.uk}$^{1}$,
Vladimir M.~Braun\footnote{Vladimir.Braun@physik.uni-regensburg.de}$^{2}$
and
Einan Gardi\footnote{Einan.Gardi@cern.ch}$^{3}$}
  \vskip0.5cm
        $^1$ {\em IPPP, Department of Physics,
University of Durham, Durham DH1 3LE, UK} \\
\vskip0.4cm
$^2$ {\em Institut f\"ur Theoretische Physik, \\ Universit\"at Regensburg,
D--93040 Regensburg, Germany}
\vskip0.4cm
$^3${\em School of Physics, The University of Edinburgh,\\
Edinburgh EH9 3JZ, Scotland, UK}


\vskip3cm

{\large\bf Abstract\\[10pt]} \parbox[t]{\textwidth}{
The QCD description of exclusive decays of the $\Lambda_b$ baryon involves hadronic matrix elements of non-local light ray operators,
the light-cone distribution amplitudes.  We introduce the complete set of three--quark distribution
amplitudes and calculate the renormalization scale dependence for the leading twist.
At leading order in the strong coupling the evolution is driven by pairwise two--quark interactions: heavy--light 
involving Sudakov logarithms as in the B-meson case, and light--light as in light mesons. 
We solve the evolution equation and show that its main effect is to generate a radiative tail extending to high energies.
Finally, we present simple models for the distribution amplitudes based on QCD sum rules, and study the effect of the evolution on these models.
}

\vfill

\end{center}
\end{titlepage}

\newpage

{\large\bf 1.}~~Heavy baryons containing a $b$-quark 
will be copiously produced at the LHC and their weak decays may provide important clues on flavour--changing currents beyond the Standard Model. 
A particular advantage of $\Lambda_b$ baryons over B mesons is their spin.
Their polarization facilitates the study of spin correlations, providing valuable information on the chirality of the short-distance transition.
This can be exploited for example at LHCb by studying rare radiative decays such as $\Lambda_b\to \Lambda \gamma$~\cite{Hiller:2007ur}.

The theory of $b$-baryon decays into light hadrons
is, however, more complicated and was receiving less attention compared to the B-meson decays. In particular, we are not aware of any dedicated study of the heavy--baryon distribution amplitudes (DAs) that are the primary non-perturbative objects required for calculating heavy--baryon decays into light particles based on the heavy quark expansion, see e.g. \cite{He:2006ud}, 
or using sum rules of the type proposed in \cite{Khodjamirian:2006st,DeFazio:2005dx,DeFazio:2007hw}.
 The only existing models 
of heavy baryon DAs \cite{Loinaz:1995wz,Hussain:1990uu} are motivated by
quark models and are not consistent with QCD constraints. 
In this letter we give, for the first time, the complete classification 
of three--quark DAs of the $\Lambda_b$ baryon in QCD in the heavy quark limit 
and discuss some of their main features. In particular we derive a renormalization--group equation that governs the scale-dependence of the leading--twist DA and study its solution. Simple models of the DAs are suggested, and their parameters are fixed based on estimates of the first few moments using QCD sum rules.

\vskip0.3cm
{\large\bf 2.}~~The $\Lambda_b$ distribution amplitudes can be defined as matrix elements of 
non-local light-ray operators built of an effective heavy quark and two light quarks following, on one hand,  the similar construction for B-mesons 
\cite{Grozin:1996pq,Kawamura:2001jm,Lange:2003ff,Braun:2003wx,Lee:2005gza,Kawamura:2008qu}
and, on the other hand,  the QCD description of nucleon DAs \cite{Chernyak:1983ej,Braun:2000kw}:
\begin{eqnarray}
\label{def:DA}
 \epsilon^{abc}\langle 0| \left(u^a(t_1 n)C\gamma_5 \nsl d^b(t_2 n)\right) h^c_{v}(0)
  |\Lambda(v)\rangle &=& f^{(2)}_\Lambda \Psi_2(t_1,t_2) \Lambda(v)\,,
\nonumber\\ 
 \epsilon^{abc}\langle 0| \left(u^a(t_1 n)C\gamma_5 d^b(t_2 n)\right) h^c_{v}(0)
  |\Lambda(v)\rangle &=& f^{(1)}_\Lambda \Psi_3^s(t_1,t_2) \Lambda(v)\,,
\nonumber\\ 
 \epsilon^{abc}\langle 0| \left(u^a(t_1 n)C\gamma_5 
 i  \sigma_{\bar n n} d^b(t_2 n)\right) h^c_{v}(0)
  |\Lambda(v)\rangle &=& 2 f^{(1)}_\Lambda \Psi_3^\sigma(t_1,t_2) \Lambda(v)\,,
\nonumber\\ 
 \epsilon^{abc}\langle 0| \left(u^a(t_1 n)C\gamma_5 \nbarsl d^b(t_2 n)\right) h^c_{v}(0)
  |\Lambda(v)\rangle &=& f^{(2)}_\Lambda \Psi_4(t_1,t_2) \Lambda(v)\,.
\end{eqnarray}
Here the subscript 2, 3, 4 refers to the twist of the diquark operator, $C$ is the charge conjugation matrix, 
$\Lambda(v)$ is the Dirac spinor, $\vsl \Lambda(v) = \Lambda(v)$. The non-relativistic normalisation is
assumed, $\bar \Lambda \Lambda =1$. Further, $n_\mu$ and $\bar n_\mu$ are light-like vectors which we 
choose such that $v_\mu = (n_\mu+\bar n_\mu)/2$, $v\cdot n=1$, $n\cdot \bar n =2$.

The DAs $\Psi_2, \Psi_3^s, \Psi_4$ are symmetric under the interchange of the coordinates of the two light quarks; $\Psi_3^\sigma$ is antisymmetric.
The couplings $f_\Lambda^{(i)}$ are given by the matrix elements of the \emph{local} operators
\begin{eqnarray}
  \epsilon^{abc}\langle 0| \left(u^a(0) C\gamma_5  d^b(0)\right) h^c_{v}(0)
  |\Lambda(v)\rangle &=& f^{(1)}_\Lambda \Lambda(v)\,,
\nonumber\\
 \epsilon^{abc}\langle 0| \left(u^a (0)C\gamma_5  \vsl d^b(0)\right) h^c_{v}(0)
  |\Lambda(v)\rangle &=& f^{(2)}_\Lambda \Lambda(v)\,.
\label{def:flambda}
\end{eqnarray} 
and are used in (\ref{def:DA}) for convenience.

Alternatively, one can define
\begin{eqnarray}
 \epsilon^{abc}\langle 0| \left(u^a(t_1 n)C\gamma_5 \nsl {\nbarsl} 
d^b(t_2 n)\right) h^c_{v}(0)
  |\Lambda(v)\rangle &=& f^{(1)}_\Lambda \Psi_3^{+-}(t_1,t_2) \Lambda(v)\,,
\nonumber\\ 
 \epsilon^{abc}\langle 0| \left(u^a(t_1 n)C\gamma_5 {\nbarsl} 
 \nsl d^b(t_2 n)\right) h^c_{v}(0)
  |\Lambda(v)\rangle &=& f^{(1)}_\Lambda \Psi_3^{-+}(t_1,t_2) \Lambda(v)\,.
\label{def:+-}
\end{eqnarray}
In contrast to the above, the DAs in (\ref{def:+-}) do not have any 
definite symmetry;
however, the isospin zero condition for the diquark implies that
\begin{equation}
  \Psi_3^{-+}(t_1,t_2) =  \Psi_3^{+-}(t_2,t_1) \,.
\end{equation}
It follows that 
\begin{eqnarray}
   \Psi^s_3(t_1,t_2) = \frac{1}{4}\Big[\Psi_3^{+-}(t_1,t_2)+ \Psi_3^{+-}(t_2,t_1)\Big]\,,
\nonumber\\
   \Psi^\sigma_3(t_1,t_2) = \frac{1}{4}\Big[\Psi_3^{+-}(t_1,t_2) - \Psi_3^{+-}(t_2,t_1)\Big]\,,
\end{eqnarray} 
correspond to the symmetric (antisymmetric) parts of $\Psi_3^{+-}$, respectively.

Going over to the momentum space, we define
\begin{eqnarray}
  \Psi(t_1,t_2) &=& \int_0^\infty \!\!d\omega_1  \int_0^\infty\!\! d\omega_2 \, e^{-it_1\omega_1 -it_2\omega_2} \psi(\omega_1,\omega_2) 
=
\int_0^\infty\! \!\omega\, d\omega \int_0^1 du \,  e^{-i\omega(t_1u +it_2\bar u)} \widetilde\psi(\omega,u)\,, 
\label{eq:norm}
\end{eqnarray}
so that 
\begin{equation}
  \widetilde\psi(\omega,u) =  \psi(u \omega,\bar u \omega) 
\end{equation}
where $\bar u = 1-u$. In the first representation $\omega_1$ and $\omega_2$ are the energies of the $u$- and $d$-quark, respectively,
and in the second one $\omega=\omega_1+\omega_2$ is the total energy carried by light quarks (in the heavy--quark 
rest frame) whereas the dimensionless variable $u$ corresponds to the energy fraction carried by the $u$-quark, i.e. $\omega_1 = u \omega$
and  $\omega_2 = \bar u \omega$. These two representations are fully equivalent and can be convenient in different contexts.  

A non-relativistic constituent quark picture of the $\Lambda_b$ suggests that 
$f^{(2)}_\Lambda \simeq f^{(1)}_\Lambda $ at low scales of order 1 GeV, and this expectation is supported by 
numerous  QCD sum rule calculations \cite{Bagan:1993ii,Grozin:1992td,Groote:1996em,Groote:1997yr}.
In fact, the difference between the two couplings is only obtained at the level of NLO perturbative
corrections to the sum rules  \cite{Groote:1996em,Groote:1997yr} and it is numerically small.

The anomalous dimensions of the operators in (\ref{def:flambda}) 
\begin{equation}
\label{f_scale_dep}
\frac{d\ln f_{\Lambda}^{(i)}(\mu)}{d\ln \mu}
\equiv -\gamma^{(i)} = - \sum_k a(\mu)^k \gamma^{(i)}_k;\qquad a(\mu)\equiv \alpha_s^{\MSbar}(\mu)/(4\pi) 
\end{equation}
are known to NLO \cite{Groote:1996em}\footnote{$\gamma^{(i)}_2$ quoted here are in the naive $\gamma_5$ scheme.}:
\begin{eqnarray}
  \gamma^{(1)}_1  =-8 &\qquad& \gamma^{(1)}_2 = -\frac{1}{9}\Big[796-16\zeta(2)-40n_f\Big]\,,
\label{gamma1}\\
  \gamma^{(2)}_1  =-4 &\qquad& \gamma^{(2)}_2 = -\frac{1}{9}\Big[322-16\zeta(2)-20n_f\Big]\,.
\label{gamma2}
\end{eqnarray}
Thus, the scale dependence of the couplings is given by 
\begin{eqnarray}
 f^{(i)}_\Lambda(\mu) &=& f^{(i)}_\Lambda(\mu_0)
 \left(\frac{\alpha_s(\mu)}{\alpha_s(\mu_0)}\right)^{\gamma^{(i)}_1/\beta_0}
 \left(1-\frac{\alpha_s(\mu_0)-\alpha_s(\mu)}{4\pi}\frac{\gamma^{(i)}_1}{\beta_0}
 \left(\frac{\gamma^{(i)}_2}{\gamma^{(i)}_1}-\frac{\beta_1}{\beta_0}\right) \right),
\end{eqnarray}
where $da(\mu)/d\ln \mu=-\beta_0 a(\mu)^2 -\beta_1 a(\mu)^3+\cdots$ with $\beta_0 = 2(11-2/3 n_f)$, $\beta_1=4(51-19/3 n_f)$.

For the numerical value of the couplings we quote the result of the NLO QCD sum rule 
analysis in Ref.~\cite{Groote:1997yr}:
\begin{equation}
\label{f_numbers}
  f^{(2)}_\Lambda \simeq f^{(1)}_\Lambda \simeq  0.030\pm 0.005~\mbox{\rm GeV}^3
\end{equation}
at the renormalization scale $\mu=1$~GeV.  Note that these couplings cannot coincide at all scales since the corresponding operators have different anomalous dimensions. 

Similarly to the B-meson case \cite{Beneke:2000wa,Kawamura:2001jm,Kawamura:2008qu}
QCD equations of motion can be used to derive exact relations between the three--quark 
DAs in (\ref{def:DA}) and the four--particle DAs involving an extra gluon field strength tensor. The corresponding analysis will be presented elsewhere.

\vskip0.3cm
{\large\bf 3.}~~The DAs in (\ref{def:DA}) are scale dependent. 
The leading-order (LO)  evolution equation for the 
leading-twist DA $\psi_2(\omega_1,\omega_2;\mu)$ can be derived following the 
usual procedure by identifying the ultraviolet singularities of 
one-gluon-exchange diagrams. 

The result can be expressed 
in terms of the two-particle kernels familiar from the evolution equations of
 the $B$-meson and $\pi$-meson DAs. We obtain
\begin{eqnarray}
  \mu \frac{d}{d\mu} \psi_2(\omega_1,\omega_2;\mu) &=&
-\frac{\alpha_s(\mu)}{2\pi}\left(1+\frac{1}{N_c}\right) \Bigg\{
   \int_0^\infty d\omega_1'\, 
     \gamma^{\rm LN}(\omega_1',\omega_1;\mu) \psi_2(\omega'_1,\omega_2;\mu)
\nonumber\\&&{}
   + \int_0^\infty d\omega_2'\, 
     \gamma^{\rm LN}(\omega_2',\omega_2;\mu) \psi_2(\omega_1,\omega'_2;\mu)
\nonumber\\&&{}
   - \int_0^1 dv\, V(u,v) 
\psi_2(v \omega, \bar v \omega;\mu)+ \frac{3}{2}\, \psi_2(\omega_1,\omega_2;\mu)\Bigg\}
\label{eq:evolution}
\end{eqnarray}
where in the last line $\omega\equiv \omega_1+\omega_2$ and 
$u\equiv\omega_1/(\omega_1+\omega_2)$; the last term in the curly brackets,
$\frac{3}{2}\,\psi_2$, is a result of the subtraction of the one-loop renormalization of the coupling $f^{(2)}_\Lambda$ according to Eqs. (\ref{f_scale_dep}) and (\ref{gamma2}).

The first two convolution integrals in Eq.~(\ref{eq:evolution}) are associated with heavy--light dynamics: each of them involves just one of the light quarks. Indeed, the kernel $\gamma^{\rm LN}(\omega',\omega;\mu)$ coincides with the one controlling the evolution of the B--meson distribution amplitude, the Lange-Neubert anomalous dimension \cite{Lange:2003ff}
\begin{eqnarray}
\label{eq:LN}
  \gamma^{\rm LN}(\omega',\omega;\mu) &=& 
     \left( \ln \frac{\mu}{\omega} -\frac{5}{4}\right)\delta(\omega-\omega')
- \Gamma_{\rm LN}(\omega',\omega)\,\nonumber \\
\Gamma_{\rm LN}(\omega',\omega)&\equiv& \left[ \frac{\omega}{\omega'}\frac{\theta(\omega'-\omega)}{\omega'-\omega}
 + \frac{\theta(\omega-\omega')}{\omega-\omega'} 
\right]_\oplus
\end{eqnarray}
where
\begin{equation}
  \int_0^\infty d\omega' \Big[\gamma(\omega',\omega)\Big]_\oplus f(\omega')
  = \int_0^\infty d\omega' \,\gamma(\omega',\omega) \,\Big[f(\omega')-f(\omega)\Big]\,.
\end{equation}

In turn, the last convolution integral in Eq.~(\ref{eq:evolution}) describes the interaction between the light quarks. $V(u,v)$ is the celebrated ER-BL kernel \cite{LB80}:
\begin{equation}
 V(u,v) = 
      \left[\frac{1-u}{1-v}\left(1+\frac{1}{u-v}\right)\theta(u-v)
   + \frac{u}{v}\left(1+\frac{1}{v-u}\right)\theta(v-u)\right]_+\, ,
\end{equation}
where the ``+'' subtraction is defined as
\begin{equation}
  [V(u,v)]_+ = V(u,v)-\delta(u-v)\int_0^1\,dt\,\, V(t,v)\,.
\end{equation}

Note that in Eq.~(\ref{eq:evolution}) we retain the dependence on the number of colors $N_c$ in the prefactor although the whole construction only makes sense for $N_c=3$. 

{\large\bf 4.}~~
{}For small evolution ranges, $\ln(\mu/\mu_0)\lsim 1$, it is sufficient to interpret the derivative on the 
l.h.s.~of (\ref{eq:evolution}) as a finite difference
$[\psi_2(\omega_1,\omega_2;\mu)-\psi_2(\omega_1,\omega_2;\mu_0)]/\ln(\mu/\mu_0)$ and 
substitute the initial condition $\psi_2(\omega_1,\omega_2;\mu_0)$ for $\psi_2(\omega_1,\omega_2;\mu)$ on the r.h.s.\footnote{Note that the scale $\mu$ appearing explicitly in (\ref{eq:LN}) and the scale of the strong coupling must be the same.}.
Obviously, this corresponds to taking into account one-loop renormalisation only, neglecting the resummation of potentially large logarithms. 
 As we shall see below (Figures \ref{fig:model_evolution_fixed_w} and \ref{fig:model_evolution_fixed_u}) this 
single--evolution--step (one--loop) approximation is quite 
good in practice, e.g. for $\mu_0=1$ GeV and $\mu\simeq m_b/2$.
 
In order to go beyond the one-loop approximation, one possibility is to integrate the evolution equation (\ref{eq:evolution})
numerically. We have taken another, semi-analytic, approach which has an advantage that it allows one 
to understand the structure of the solution. 
To this end we first remove the $\ln(\mu)$ term on the r.h.s, which is related to the cusp anomalous dimension, by defining:
\begin{equation}
\label{phi_def}
\psi_2(\omega_1,\omega_2;\mu)=\phi(\omega_1,\omega_2;\mu)\,\left( \frac{\omega_1\omega_2}{\mu^2}\right)^{g(\alpha_s(\mu))/2},
\end{equation}
where 
\begin{equation}
g(\alpha_s(\mu))=\int_{\mu_{0}}^{\mu} \frac{dm}{m}
\Gamma_{\cusp}(\alpha_s(m)),\qquad \Gamma_{\cusp}(\alpha_s(m))=\frac{C_F\alpha_s(m)}{4\pi}+\cdots\,.
\end{equation}
Substituting (\ref{phi_def}) in  (\ref{eq:evolution}) yields an evolution equation for $\phi(\omega_1,\omega_2;\mu)$. The next step is to 
go over to the moments space:
\begin{equation}
\tilde{\phi}(N,M;\mu)=\int_0^{\infty} d\omega_1 \omega_1^{N-1}
\int_0^{\infty} d\omega_2 \omega_2^{M-1}
\,\phi(\omega_1,\omega_2;\mu)\,.
\end{equation}
This leads to factorization of the Lange--Neubert  terms
since
\begin{align}
\int_0^{\infty}d\omega \omega^{N-1} \Gamma_{\rm LN}(\omega',\omega) =(\omega')^{N-1} \tilde{\Gamma}_{\rm LN}(N);\qquad
 \tilde{\Gamma}_{\rm LN}(N)&=-\Psi(N)-\Psi(-N)-2\gamma_E\,\,.
\end{align}
The ER-BL term does not factorize\footnote{We use the shorthand notation $\bar{u}\equiv 1-u$, $\bar{v}\equiv 1-v$.}:  
\begin{align}
\label{Gamma_V_def}
\begin{split}
\tilde{\Gamma}_V^{\phi}(N,M;\mu)&\equiv 
\displaystyle{\frac{\displaystyle{\int_0^1 dv \int_0^1 du \int_{0}^{\infty}d\omega \omega
(u\omega)^{N-1} (\bar{u}\omega)^{M-1}  
\left(\frac{v\bar{v}}{u\bar{u}}\right)^{g(\alpha_s(\mu))/2} V(u,v) \,\,\phi(v\omega,\bar{v}\omega;\mu)}}{\tilde{\phi}(N,M;\mu)}}\,,
\end{split}
\end{align}
calling for some approximation. A simple one is obtained by substituting the initial condition $\phi(v\omega, \bar{v}\omega; \mu_0)$ for $\phi(v\omega, \bar{v}\omega; \mu)$ in both the numerator and denominator of (\ref{Gamma_V_def}). While this can be a starting point for an iterative procedure, we find that in practice such iteration is not necessary owing to the smallness of the ER-BL term. 

With this assumption, the r.h.s.~of the evolution equation for $\tilde{\phi}(N,M;\mu)$ factorizes
 leading to exponentiation of the kernels: 
\begin{align}
\begin{split}
\tilde{\phi}(N,M;\mu)=\tilde{\phi}(N,M;{\mu}_0)
\,\exp\left\{
\int_{\mu_0}^{\mu}\frac{dm}{m}\, E(N,M,m)
\right\}
\end{split}
\end{align}
where the exponent is given by:
\begin{align}
\begin{split}
E(N,M,m)&=g(\alpha_s(m))+\frac{2\alpha_s(m)}{3\pi}\,
\Bigg[1 +\tilde{\Gamma}_{\text{LN}}\Big(N-g(\alpha_s(m))/2\Big) \\&
+\tilde{\Gamma}_{\text{LN}}\Big(M-g(\alpha_s(m))/2\Big) 
+\tilde{\Gamma}_V^{\phi}(N,M;m) 
\Bigg]\,,
\end{split}
\end{align}
resumming double-- as well as single--log terms to all orders.

Further simplification in evaluating $\tilde{\Gamma}_V^{\phi}(N,M;\mu)$ is achieved by replacing the ER-BL kernel $V(u,v)$ in (\ref{Gamma_V_def}) by the expansion as a sum of products of Gegenbauer polynomials (see e.g.~\cite{Braun:2003rp}):
\begin{equation}
 V(u,v) = -u(1-u)\sum_{n=0}^\infty\frac{2(2n+3)}{(n+1)(n+2)}\gamma_n C^{3/2}_n(2u-1)C^{3/2}_n(2v-1)\,,
\label{V-expand}
\end{equation}
where $\gamma_n =  1-2/[(n+1)(n+2)] + 4\sum_{m=2}^{n+1}1/m$ 
is the leading-order anomalous dimension. In the numerical evaluation presented below we truncate the sum in (\ref{V-expand}) at the leading non-trivial term, $n=2$. The impact of this truncation proves to be small, at least for the models of the DA that we consider.

Finally, the answer in the momentum (energy) space is restored by a double inverse--Mellin transform which is done numerically. 
The main effect of the evolution is to generate a `radiative tail' of the DA 
that falls off as $\ln(\omega_1/\mu)/\omega_1$ or $\ln(\omega_2/\mu)/\omega_2$ at large energies, which is the same effect that the evolution has on the B-meson DA, see \cite{Lange:2003ff,Braun:2003wx,Lee:2005gza,Kawamura:2008qu}.

\vskip0.3cm
{\large\bf 5.}~~Realistic models for the DAs can be obtained using QCD sum rules for the 
correlation functions involving the non-local light-ray operators in (\ref{def:DA}) and a 
suitable local current.
We define
\begin{equation}
 \bar J(x) =  \epsilon^{abc} \left(\bar d^a(x) P_+\gamma_5 C^T \bar u^b(x)\right)\bar h_v^c(x)
\label{current}
\end{equation}
where $P_+ = (1+\,\vsl)/2$ and consider, for the leading twist,  the correlation function
\begin{eqnarray}
\lefteqn{ \hspace*{-4cm}i\int d^4 x\, e^{-i E vx} 
 \epsilon^{abc} \langle 0| (u^a(t_1 n)C\gamma_5 \nsl d^b(t_2 n)) h^c_{v}(0)  \bar J(x)|0\rangle =}
\nonumber\\\hspace*{3cm}&=&  P_+ \int_0^\infty \omega d\omega \int_0^1 du\, e^{-i\omega(ut_1+\bar u t_2)} \Pi_2(\omega,u;E)
\end{eqnarray}
and similarly for the other structures.  The general form  of the sum rule is then
\begin{eqnarray}
\frac{1}{2}|f_\Lambda^{(2)}|^2 \widetilde \psi^{SR}_2(\omega,u) e^{-\bar\Lambda/\tau}&=&
 {\mathbb B}[\Pi_2](\omega,u;\tau,s_0)\,,
\end{eqnarray}
where ${\mathbb B}[\Pi_2](\omega,u;\tau,s_0)$ is the Borel-transformed continuum-subtracted 
invariant function $\Pi_2(\omega,u;E)$; $\tau$ is the Borel parameter which we take to be in the interval 
$0.4 < \tau < 0.8$~GeV and $s_0=1.2$~GeV is the continuum threshold (interval of duality);
$\bar\Lambda = m_{\Lambda_b}-m_b \simeq 0.8$~GeV. 

Taking into account only the leading-order perturbative contribution to the sum rule, one obtains 
\begin{eqnarray}
\widetilde \psi_2(\omega,u) &=& \frac{15}{2} {\cal N}^{-1} \omega^2 \bar u u \int_{\omega/2}^{s_0} ds\, e^{-s/\tau} (s-\omega/2)\,,
\nonumber\\
\widetilde \psi_4(\omega,u) &=& 5 {\cal N}^{-1} \int_{\omega/2}^{s_0} ds\, e^{-s/\tau} (s-\omega/2)^3\,,
\nonumber\\
\widetilde \psi_3^s(\omega,u) &=& \frac{15}{4} {\cal N}^{-1} \omega  \int_{\omega/2}^{s_0} ds\, e^{-s/\tau} (s-\omega/2)^2\,,
\nonumber\\
\widetilde \psi_3^\sigma(\omega,u) &=& \frac{15}{4} {\cal N}^{-1} \omega (2u-1) \int_{\omega/2}^{s_0} ds\, e^{-s/\tau} (s-\omega/2)^2\,,
\label{SR:pert}
\end{eqnarray}
with 
\begin{eqnarray}
   {\cal N} &=& \int_0^{s_0}ds\, s^5 e^{-s/\tau}\,. 
\end{eqnarray}
To this accuracy the coupling is equal to 
$|f_\Lambda|^2 = e^{\bar\Lambda/\tau}\mathcal{N}/(20\pi^4)$.
All DAs have in this approximation the support property 
$0<\omega<2s_0$ and are normalized such that
\begin{equation}
 \int_0^{2s_0}\,\,\omega d\omega\int_0^1 du\, \widetilde\psi_2(\omega,u) =
  \int_0^{2s_0}\,\,\omega d\omega\int_0^1 du\, \widetilde\psi_3^s(\omega,u) =
 \int_0^{2s_0}\,\,\omega d\omega\int_0^1 du\,  \widetilde\psi_4(\omega,u) = 1\,. 
\end{equation}
Note that the leading--twist DA $\widetilde\psi_2(\omega,u) \equiv  \psi_2(\omega_1,\omega_2)$ vanishes when either
one of the light-quark energies goes to zero: $\omega^2 u (1-u) = \omega_1\omega_2$.  This property
is model-independent and consistent with the evolution equation in (\ref{eq:evolution}).

The limit $\tau\to\infty$ is known as the approximation of local duality. In this case one obtains, 
for example
\begin{equation}
\psi_2^{\rm LD}(\omega_1,\omega_2) = \frac{45}{8s_0^6}\omega_1\omega_2 (2s_0-\omega_1-\omega_2)^2\theta(2s_0-\omega_1-\omega_2)
\label{SR:LD}
\end{equation}
and similarly for other twists. 
With decreasing Borel parameter the DA becomes tilted towards smaller momenta since contributions of large $\omega \to 2 s_0$
are more strongly affected by the additional (exponential) suppression factor. 

{}As explained in \cite{Grozin:1996pq,Braun:2003wx} in order to evaluate the 
non-perturbative contributions to the sum rule one is forced to use the non-local condensates. We use the general 
parametrisation~\cite{Mikhailov:1986be,Mikhailov:1991pt}
\begin{eqnarray}
   \langle \bar q(x) q(0)\rangle &=& \langle \bar q q\rangle \int_0^\infty d\nu\, e^{\nu x^2/4}f(\nu)  
\end{eqnarray} 
where $\langle \bar q q\rangle \simeq -(240$~MeV$)^3$ is the quark condensate and 
$f(\nu)$ is  the model function \cite{Braun:1994jq,Braun:2003wx}
\begin{eqnarray}
   f(\nu) &=& \frac{\lambda^{a-2}}{\Gamma(a-2)} \nu^{1-a} e^{-\lambda/\nu};\quad a-3 = 4\frac{\lambda}{m_0^2}\,;
\end{eqnarray} 
$m_0^2 \simeq 0.8$~GeV$^2$ is the standard notation for the ratio of the mixed quark-gluon and quark condensates,
and $\lambda\simeq (400$~MeV$)^2$ is the correlation length.

Using this model, we obtain the sum rule:
\begin{eqnarray}
\frac{1}{2}|f_\Lambda^{(2)}|^2 \widetilde \psi^{SR}_2(\omega,u) e^{-\bar\Lambda/\tau}&=&
\frac{3}{16\pi^4} \omega^2 \bar u u \int_{\omega/2}^{s_0} ds\, e^{-s/\tau} (s-\omega/2)
\nonumber\\
&&{}-\frac{\langle \bar u u\rangle}{8\pi^2}\frac{\bar u}{u}\, \kappa^{a-2}
 \frac{\sin a\pi}{\pi(3-a)}
\Bigg[ \left(s_0-\omega/2-\kappa\right)^{3-a}e^{-s_0/\tau}
\nonumber\\
&&{}
 +\frac{1}{\tau}\int_{\omega/2+\kappa}^{s_0}ds\,
 e^{-s/\tau}\left(s-\omega/2-\kappa\right)^{3-a}\Bigg]
\nonumber\\
&&{}-\frac{\langle \bar d d\rangle}{8\pi^2}\frac{u}{\bar u}\, {\bar\kappa}^{a-2}
 \frac{\sin a\pi}{\pi(3-a)}
\Bigg[ \left(s_0-\omega/2-\bar\kappa\right)^{3-a} e^{-s_0/\tau}
\nonumber\\
&&{}
 +\frac{1}{\tau}\int_{\omega/2+\bar\kappa}^{s_0}
ds\,e^{-s/\tau}\left(s-\omega/2 -\bar\kappa)\right)^{3-a}\Bigg]
\nonumber\\
&&{}+\frac13 \langle \bar u u\rangle \langle \bar d d\rangle \tau^2 f(2\tau u \omega) f(2\tau \bar u \omega) e^{-\omega/(2\tau)}
\label{SR:full}
\end{eqnarray} 
where Heaviside functions of the difference between the upper and the lower limits of integration are implied, and where we used a shorthand notation
\begin{eqnarray}
 \kappa =  \frac{\lambda}{ 2u\omega}, \qquad
 \bar \kappa = \frac{\lambda}{2\bar u\omega}\,.
\end{eqnarray}

\begin{figure}[t]
\centerline{
\begin{picture}(200,90)(0,0)
\put(50,0){\epsfxsize8cm\epsffile{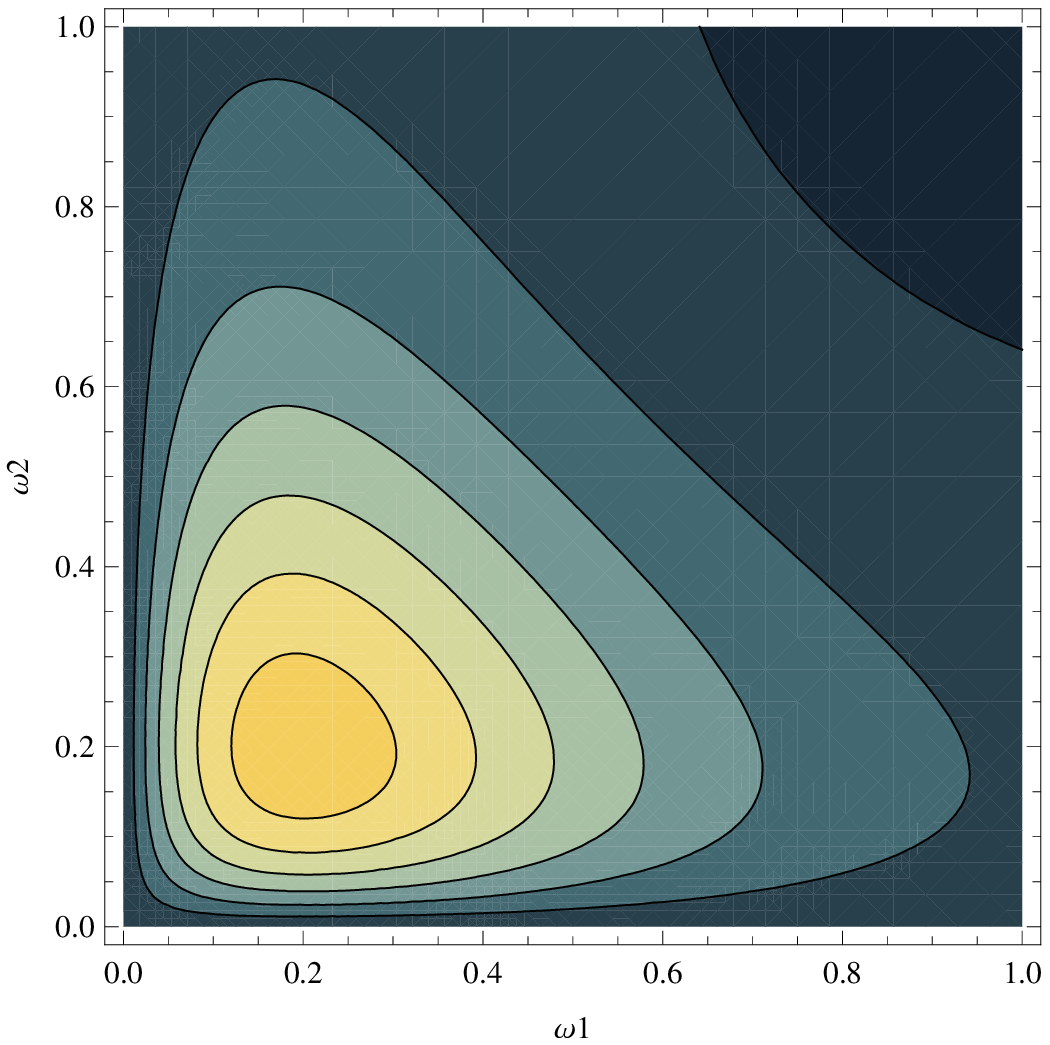}}
\put(135,13){\epsfxsize1.5cm\epsffile{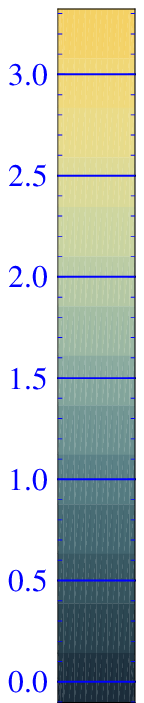}}
\end{picture}
}
\caption{\small
QCD model for the leading twist DA of the $\Lambda_b$ baryon defined in Eq.~(\ref{model1})
}
\label{fig:model}
\end{figure}


\begin{figure}[htb]
\centerline{\epsfig{file=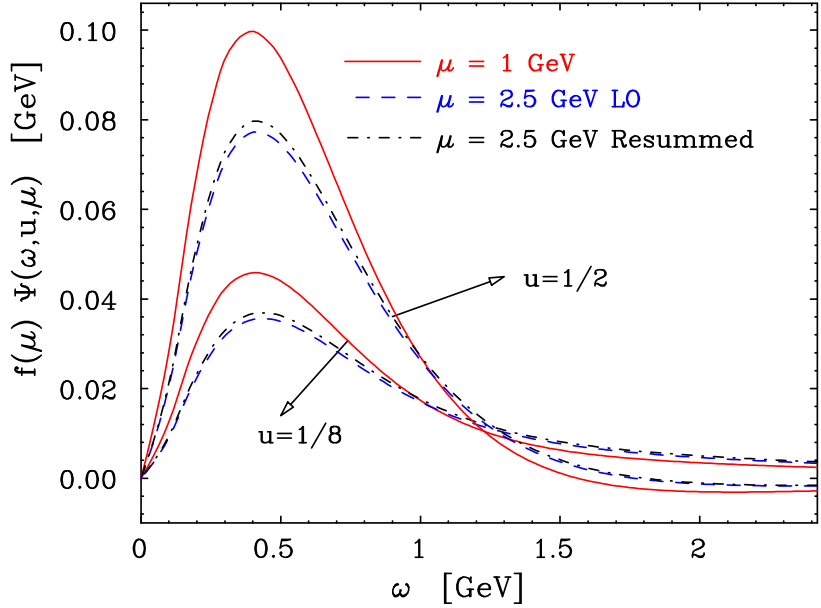, height=8.0cm}}
\caption{\small
QCD model for the leading--twist DA of the $\Lambda_b$ baryon defined in Eq.~(\ref{model1}) at the 
scale of 1 GeV (solid curve) and after the evolution to $\mu=2.5$~GeV (dash-dotted curve)  as a function of 
$\omega=\omega_1+\omega_2$ for two values of the light quark momentum fraction 
$u=0.5$ and $u=0.125$. The result of a single--step evolution to $\mu=2.5$~GeV, which includes the $\sim {\cal O}(\alpha_s)$ correction only, is shown by dashes for comparison. 
}
\label{fig:model_evolution_fixed_w}
\end{figure}

\begin{figure}[htb]
\centerline{\epsfig{file=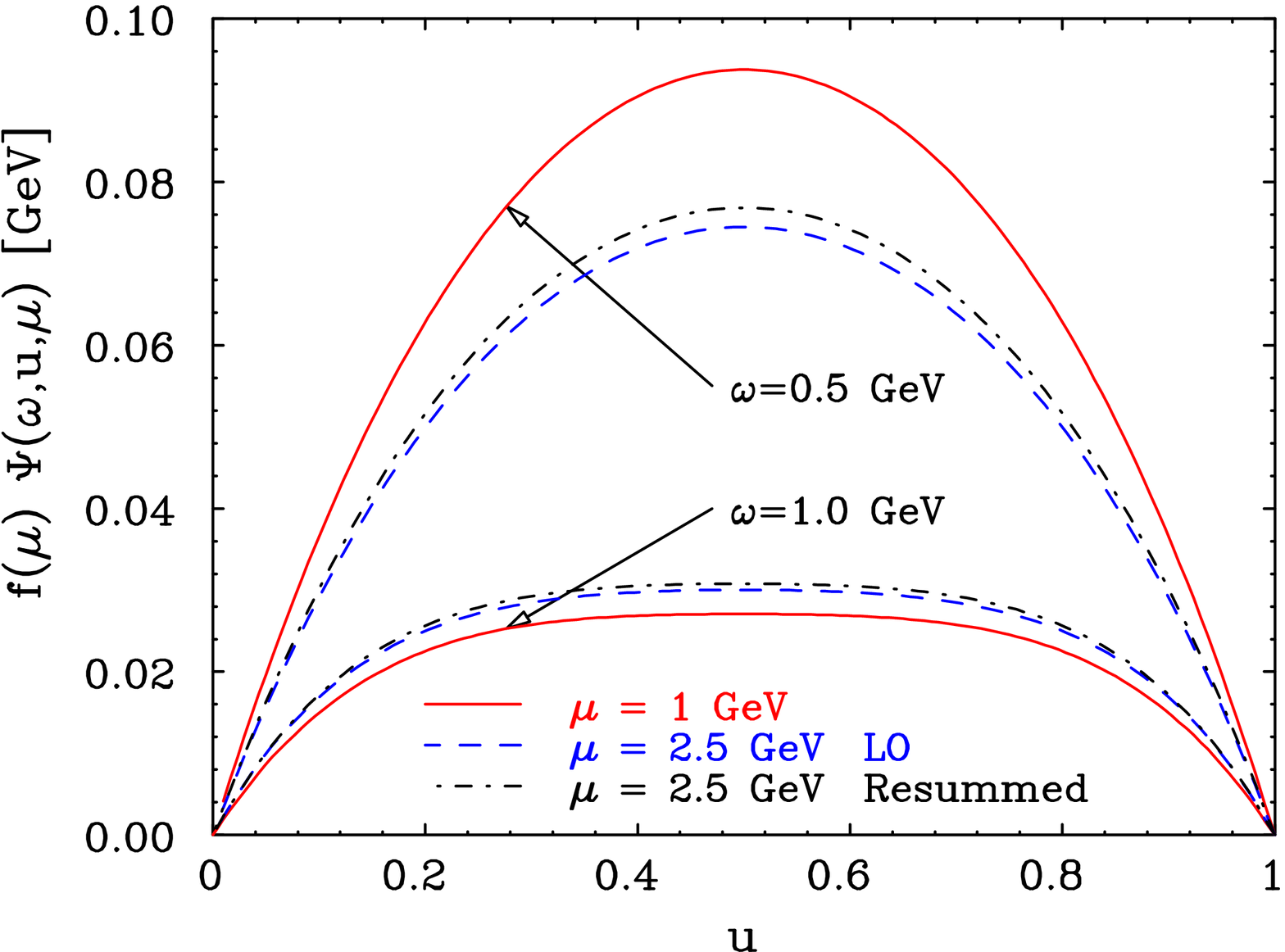, height=8.0cm}}
\caption{\small
The $u$ dependence of the DA for fixed $\omega=0.5$~GeV (near the peak in Figure~\ref{fig:model_evolution_fixed_w}) and $\omega=1.0$~GeV (crossing over to the tail region). The curves are as explained in Fig.~2. Note that the effect of evolution to higher $\mu$ is to decrease the DA for any $u$ in the former case and increase it in the latter.
}
\label{fig:model_evolution_fixed_u}
\end{figure}

{}From the vast experience of QCD sum rule calculations of the pion DA 
(see e.g. \cite{Chernyak:1981zz,Mikhailov:1986be,Braun:1988qv,Bakulev:2001pa})
it is known, however, that the QCD sum rules
cannot give the functional form of the DAs but rather have to be used to constrain certain momentum fraction integrals 
(the moments). 
Furthermore, obtaining a meaningful error estimate is especially difficult in the present case 
because there is not enough experience in using the concept of non-local 
condensates in baryon sum rules. In order to be on the conservative side we adopt the following procedure.
As well known, QCD sum rules can be written for different interpolating currents. 
Our choice in (\ref{current}) corresponds to the constituent type sum rule, in the terminology of 
Ref.~\cite{Groote:1997yr}, and has the advantage that the corresponding sum rule (\ref{SR:full}) 
has several terms.  Replacing the projector  $P_+$ in (\ref{current}) by the unity matrix
or by $\vsl$ one obtains two other currents, called $\bar J_1$ and $\bar J_2$  in \cite{Groote:1997yr}. 
Obviously $\bar J=(\bar J_1+\bar J_2)/2$. The corresponding sum rules pick up contributions 
of even and odd dimension in (\ref{SR:full}) respectively, i.e.  perturbation theory and the quartic condensate for $\bar J_2$
and the quark condensate for $\bar J_1$. We take the sum rule in (\ref{SR:full}) for 
our central values and use the spread of the results using  $\bar J_1$ and $\bar J_2$ 
(using the central values of the parameters) as an error estimate:
\begin{align}
\label{final:numb}
\begin{split}
\begin{array}{ll}
{\displaystyle \int_0^{2s_0} \!\omega d\omega\int_0^1\,\, du\, \widetilde\psi_2(\omega,u) \,\equiv\, 1\,},
&{\displaystyle  \quad \int_0^{2s_0}\! \omega d\omega\int_0^1 du\,C^{3/2}_2(2u-1)\, \widetilde\psi_2(\omega,u) \,=\, 1.0^{+0.5}_{-1.0}\,},
\\[5mm]
{\displaystyle \int_0^{2s_0} \! d\omega\int_0^1\,\, du\, \widetilde\psi_2(\omega,u) \,=\, 1.7\pm0.7\,},
&
{\displaystyle \quad
\int_0^{2s_0}\! d\omega\int_0^1 \!du\, C^{3/2}_2(2u-1)\ \widetilde\psi_2(\omega,u) = 0.6^{+0.7}_{-1.4}}\,.
\end{array}
\end{split}
\end{align}
The error bands given in (\ref{final:numb}) should be regarded as most conservative: using the sum rule (\ref{SR:full}) alone and varying the parameters in a reasonable range yields much smaller variations. In  particular the value of the first integral in the second line
in (\ref{final:numb}) is very stable with respect to variations of the Borel parameter.

Note that the ratio of the integrals with and without the $\omega$ factor is different for the Gegenbauer moment in the energy fraction as compared to the unit weight. This implies that the $\omega$-dependence of these components is different. 
Taking into account the expected 
low-energy behaviour $\sim \omega_1\omega_2$ and the sum rule moments of Eq.~(\ref{final:numb}), we propose a simple model (see Fig.~1) for the leading--twist DA at the low scale of $\mu=1$~GeV:
%
\begin{eqnarray}
\widetilde\psi_2(\omega,u) &=& \omega^2 u(1-u) \left[ \frac{1}{\varepsilon_0^4}e^{-\omega/\varepsilon_0} + 
a_2C_2^{3/2}(2u-1)  \frac{1}{\varepsilon_1^4} e^{-\omega/\varepsilon_1}
\right]
\label{model1}
\end{eqnarray}
with $\varepsilon_0= 200^{+130}_{-60}$~MeV, $\varepsilon_1= 650^{+650}_{-300}$~MeV and $a_2=0.333^{+0.250}_{-0.333}$. 

In the calculations of $\Lambda_b$ decays into light quarks using QCD factorisation one expects that integrals involving negative powers of the quark momenta will contribute, for example:
\begin{align}
\label{Lambda-dq}
\begin{split}
 \Lambda_q (\mu,\Lambda_{\rm UV})&\equiv \int_0^{\Lambda_{\rm UV}}\!\! d\omega\,\!\int_0^1 \frac{du}{u}\, \widetilde\psi_2(\omega,u;\mu) \,=\,
\int_0^\infty  d\omega_1 \!\! \int_0^\infty  \!\! d \omega_2 \, \frac{\theta(\Lambda_{\rm UV}-\omega_1-\omega_2)}{\omega_1} \psi_2(\omega_1,\omega_2;\mu)\,,
\\
 \Lambda_d (\mu,\Lambda_{\rm UV})&\equiv \int_0^{\Lambda_{\rm UV}}\!\!  d\omega \!\int_0^1 du\, \widetilde\psi_2(\omega,u;\mu) \,=\,
\int_0^\infty \!\! d\omega_1 \, \int_0^\infty \!\! d \omega_2 \, \frac{\theta(\Lambda_{\rm UV}-\omega_1-\omega_2)}{\omega_1+\omega_2} \psi_2(\omega_1,\omega_2;\mu)\,,
\end{split}
\end{align}
where an additional energy cutoff $\omega<\Lambda_{\rm UV}$ is introduced in the definition of the moments. This guarantees that the moments are
finite in presence of a radiative tail --- 
such a tail will be generated by evolution to a higher scale, even if it is not introduced at the low scale in a given initial--condition model. We recall that in the B-meson case the radiative tail $\sim \,\ln (\omega/\mu)/\omega$ renders such 
an energy cutoff necessary for any \emph{positive} moment, while the first negative moments analogous to (\ref{Lambda-dq}) would be finite in its absence. The two-dimensional integration in the $\Lambda_b$ case implies greater sensitivity of the cutoff, as even the first inverse moment $\Lambda_q$ diverges in its absence. 

For the above model without an energy cutoff one obtains:
\begin{eqnarray}
\label{Lambda_qd_values}
  \Lambda_q (1 \,{\rm GeV}) &=\displaystyle{ \frac{1}{\varepsilon_0}+\frac{a_2}{\varepsilon_1} }\simeq 5.5^{+2.5}_{-0.5}~\mbox{\rm GeV}^{-1}; \quad
 \Lambda_d (1 \,{\rm GeV})  &= \frac{1}{3\varepsilon_0} \simeq 1.7\pm 0.7~\mbox{\rm GeV}^{-1}\,.
\end{eqnarray}
Since $\varepsilon_1\gg \varepsilon_0$, the contribution of the Gegenbauer correction to the first integral is small so that a 100\% uncertainty in $a_2$ does not play a significant role.
The effect of an energy cutoff $\Lambda_{\rm UV}$ of order 2-3 GeV is already small. The central values
of $\Lambda_{d}$ and $\Lambda_q$  for  $\Lambda_{\rm UV}=2.5$ GeV are summarized in Table~\ref{tab:a1K}, where we also present the renormalization scale dependence of these moments. The effect of evolution from $\mu=\mu_0=1$ GeV to 
$\mu=2.5$ GeV on the functional form of the DA is illustrated in Figures~2 and~3. 

\begin{table}[h]
\renewcommand{\arraystretch}{1.3}
\begin{center}
\begin{tabular}{|c|c|c|c|}\hline
& $f^{(2)}_\Lambda$ [GeV$^3]$ & $\Lambda_d$ [GeV$^{-1}$]  & $\Lambda_q$ [GeV$^{-1}$]    \\ \hline
   $\mu=1$~GeV       & 0.0300       &    1.66 &  5.38  \\ 
   $\mu=1.5$~GeV     & 0.0314       &    1.52 &  4.94  \\ 
   $\mu=2.0$~GeV     & 0.0322       &    1.40 &  4.61  \\ 
   $\mu=2.5$~GeV     & 0.0329       &    1.31 &  4.34  \\ \hline  
\end{tabular}
\end{center}
\caption[]{\small  The decay constant $f^{(2)}_\Lambda$  of Eq.~(\ref{def:flambda})
and the typical integrals $\Lambda_{d}, \Lambda_q$ of Eq.~(\ref{Lambda-dq}) at different renormalization scales $\mu$. The moments are all computed at a fixed energy cutoff $\Lambda_{\rm UV}$ = 2.5 GeV.
These numbers correspond to the central values of the model described above --- the theoretical uncertainty is as 
quoted in Eqs.~(\ref{f_numbers}) and~(\ref{Lambda_qd_values}). 
}
\label{tab:a1K}
\renewcommand{\arraystretch}{1.0}
\end{table}

A similar analysis can be done for the twist-three DAs. Without going into details we present the simplest 
models that are consistent with the QCD sum rule constraints:
\begin{eqnarray}
 \widetilde\psi^s_3(\omega,u) &= \displaystyle{\frac{\omega}{2\varepsilon_3^3}}\,e ^{-\omega/\varepsilon_3}\,;
\qquad
\widetilde\psi^\sigma_3(\omega,u) &= \frac{\omega}{2\varepsilon_3^3}(2u-1)\,e ^{-\omega/\varepsilon_3}\,
\end{eqnarray}
with $\varepsilon_3 = 230$~MeV.

\vskip0.3cm
{\bf Acknowledgements:}~~
V.B. is grateful to IPPP for hospitality and financial support during his stay at Durham
University where this work was started.

\end{document}